\documentclass{article}

\usepackage{microtype}
\usepackage{graphicx}
\usepackage{subfigure}
\usepackage{booktabs} 
\usepackage{enumitem}

\usepackage{hyperref}

\usepackage[accepted]{icml2025}

\usepackage{amsmath}
\usepackage{amssymb}
\usepackage{mathtools}
\usepackage{amsthm}

\usepackage[capitalize,noabbrev]{cleveref}

\theoremstyle{plain}

\theoremstyle{definition}

\theoremstyle{remark}

\usepackage[textsize=tiny]{todonotes}

\begin{document}

\twocolumn[
\icmltitle{HELIX: Scaling Raw Audio Understanding with Hybrid Mamba-Attention Beyond the Quadratic Limit}



\icmlsetsymbol{equal}{*}

\begin{icmlauthorlist}
\icmlauthor{Khushiyant}{equal,comp}
\icmlauthor{Param Thakkar}{equal,comp_1}
\end{icmlauthorlist}

\icmlaffiliation{comp}{Department of Computer Science, University of Freiburg, Freiburg im Breisgau, Germany}
\icmlaffiliation{comp_1}{Department of Computer Engineering and Information Technology, Veermata Jijabai Technological Institute, Mumbai, India}

\icmlcorrespondingauthor{Khushiyant}{khushiyant.khushiyant@uni-freiburg.de}
\icmlcorrespondingauthor{Param Thakkar}{puthakkar\_b22@ce.vjti.ac.in}

\icmlkeywords{Machine Learning, ICML}

\vskip 0.3in
]



\printAffiliationsAndNotice{\icmlEqualContribution}

\begin{abstract}
Audio representation learning typically evaluates design choices such as input frontend, sequence backbone, and sequence length in isolation. We show that these axes are coupled, and conclusions from one setting often do not transfer to others. We introduce HELIX, a controlled framework comparing pure Mamba, pure attention, and a minimal hybrid with a single attention bottleneck. All models are parameter-matched at about 8.3M parameters to isolate architectural effects. Across six datasets, we find that the preferred input representation depends on the backbone, and that attention hurts performance on short, stationary audio but becomes important at longer sequence lengths. On a 5-minute speaker identification task with 30,000 tokens, pure attention fails with out-of-memory errors, while HELIX closes an 11.5-point gap over pure Mamba.
\end{abstract}

\section{Introduction}
Audio understanding underpins tasks ranging from environmental sound recognition\cite{10.1145/2733373.2806390} and keyword spotting\cite{warden2018speechcommandsdatasetlimitedvocabulary} to speaker identification\cite{Nagrani_2017} and long-form multilingual speech analysis\cite{7178964,wang2021voxpopulilargescalemultilingualspeech}. The dominant recipe is to convert waveforms into mel-spectrograms or MFCCs and then apply a discriminative backbone, from CNNs\cite{hershey2017cnnarchitectureslargescaleaudio} to spectrogram transformers such as AST\cite{gong2021astaudiospectrogramtransformer}. This works well enough in practice, but it throws away information. Phase is gone. The short-time Fourier window introduces its own latency and resolution trade-offs, and very long recordings remain difficult for quadratic-attention backbones regardless.

Structured State Space Models (SSMs) offer an appealing alternative. Starting with S4\cite{gu2022efficientlymodelinglongsequences} and strengthened by Mamba\cite{gu2024mambalineartimesequencemodeling}, they provide linear-time sequence modeling with input-dependent dynamics, making them well suited to long raw audio streams. But how should these components be combined? The design space remains poorly understood, in part because prior work has largely studied three decisions in isolation: raw waveforms versus spectrograms, SSMs versus transformers, and pure versus hybrid backbones. In practice these choices are coupled. A raw frontend increases token count dramatically, which changes the cost of attention in ways that interact with the backbone. A spectrogram shortens the sequence but removes acoustic detail. And a small amount of attention may matter only when information must travel far through time, something that depends on both the frontend and the task.

We study this interaction with HELIX, a controlled experimental framework that jointly varies input representation, backbone family, and attention ratio while keeping model capacity fixed. The HELIX hybrid itself combines a raw waveform frontend with a bidirectional Mamba backbone and a single global attention bottleneck; we compare it against parameter-matched pure Mamba and pure attention baselines under identical training and evaluation settings. The hybrid design is intentionally minimal: one attention layer is the smallest possible departure from a pure SSM backbone that still provides global interaction. This makes it both a practical architecture and a controlled experimental tool: by spanning the extremes of the design space (zero, one, and six attention layers) at matched capacity, we can ask not just which model wins but under what conditions. Our initial expectation was that one configuration would dominate; instead, the ranking shifts with almost every change in task or sequence length. Concretely, we contribute:
\begin{enumerate}[leftmargin=*,itemsep=2pt,topsep=4pt]
    \item A parameter-matched comparison showing that the preferred input representation depends on which backbone you pair it with. Raw-versus-spectrogram depends on the backbone and should not be evaluated in isolation.
    \item Evidence that attention is not uniformly helpful: it hurts on short stationary audio, helps on temporally structured speech, and becomes critical at long sequence lengths.
    \item Scaling to 30,000-token (5-minute) raw waveform classification. Pure attention cannot run at this length. Pure Mamba can, but leaves an 11.5-point gap that a single attention bottleneck closes.
\end{enumerate}

\section{Related Work}
Prior work on audio modeling has usually optimized one design choice at a time. Since we argue that input representation, backbone family, and sequence length are tightly coupled, we organize the literature along those same axes to make the missing comparison visible.

\paragraph{Spectrogram-first audio modeling.}
Most supervised audio classification pipelines begin by converting the waveform into a compact time-frequency representation. CNN-based systems established mel-spectrograms as a reliable default for environmental sound and large-scale audio tagging\cite{hershey2017cnnarchitectureslargescaleaudio}, and AST/SSAST showed that transformer-style patch modeling is competitive once audio has already been compressed into a short spectrogram sequence\cite{gong2021astaudiospectrogramtransformer,gong2022ssast}. These works show that strong performance is possible after aggressive frontend compression. But that compression also hides the trade-off at the center of our work: you get shorter sequences that attention can handle, at the cost of losing phase and fine temporal detail. Whether that trade-off is worth it turns out to depend on more than just the frontend.

\paragraph{Raw-waveform modeling.}
Another line of work retains the waveform and lets the model learn the relevant acoustic decomposition. Learnable filterbanks such as SincNet\cite{ravanelli2018sincnet} and LEAF\cite{zeghidour2021leaf}, together with self-supervised speech models such as wav2vec~2.0\cite{baevski2020wav2vec20frameworkselfsupervised}, HuBERT\cite{hsu2021hubertselfsupervisedspeechrepresentation}, and WavLM\cite{Chen_2022}, show that raw audio can preserve discriminative low-level cues that handcrafted frontends may suppress. However, these studies are usually not designed to answer whether the gains come from the representation itself, from scale and pretraining, or from the backbone used after tokenization. They establish that raw audio is viable, but the question of \emph{when} it is actually preferable, under a fixed parameter budget and a specific backbone, remains largely unaddressed.

\paragraph{SSMs and hybrid backbones.}
Structured State Space Models have recently become attractive for long-sequence modeling because they offer linear-time recurrence while remaining expressive enough for modern deep learning workloads\cite{gu2022efficientlymodelinglongsequences,gu2024mambalineartimesequencemodeling}. Audio-focused variants such as Audio Mamba\cite{erol2024audiomambabidirectionalstate} and SSAMBA\cite{Shams_2024} adapt this idea to classification and generation, while hybrid architectures such as Jamba\cite{lieber2024jambahybridtransformermambalanguage} suggest that a small amount of attention can complement an SSM backbone. Yet these papers typically evaluate one architectural recipe at a time. It is usually unclear whether the reported gains come from the SSM, from the addition of limited attention, from the frontend choice, or simply from operating at a sequence length where pure attention was already at a disadvantage.

What is missing is not any particular model. It is a controlled comparison where model capacity is held fixed while frontend, backbone, and attention ratio all vary, across tasks that differ in temporal structure and sequence length. That is what we set out to provide. HELIX exists to anchor this comparison, not to claim novelty as an architecture.

\section{Methodology}

The gap identified above calls for a framework that can vary frontend, backbone, and attention ratio independently while controlling for capacity. Below we describe each axis, how we match parameters across them, and the shared training protocol.

\subsection{Design Axes}

\paragraph{Input representation.}
Given audio $\mathbf{s}\in\mathbb{R}^{T}$ at 16\,kHz, we tokenize it via one of two frontends.

The \textbf{raw waveform} path applies a 1-D convolution with kernel and stride of 160 samples (10\,ms):
\begin{equation}
\mathbf{Z} = \mathrm{LayerNorm}\!\bigl(\mathrm{Conv1d}(\mathbf{s};\,k{=}160,\,s{=}160)\bigr) \;\in\;\mathbb{R}^{L\times d},
\label{eq:raw-embed}
\end{equation}
where $L=\lfloor T/160\rfloor$. A 5-second clip yields $L{=}500$; a 5-minute clip yields $L{=}30{,}000$. This path preserves phase and forces the backbone to learn its own time-frequency decomposition.

The \textbf{mel-spectrogram} path computes a 128-bin log-mel spectrogram (FFT 1024, hop 512), yielding an image $\mathbf{M}\in\mathbb{R}^{1\times128\times F}$, then applies a 2-D convolution with $16{\times}16$ kernel and stride:
\begin{equation}
\mathbf{Z} = \mathrm{LayerNorm}\!\bigl(\mathrm{Reshape}\!\bigl(\mathrm{Conv2d}(\mathbf{M})\bigr)\bigr) \;\in\;\mathbb{R}^{L'\times d},
\label{eq:spec-embed}
\end{equation}
where $L'=(128/16)\times\lceil F/16\rceil \approx 80$ for the same 5-second clip. The spectrogram path produces over $6\times$ fewer tokens than the raw path. This matters more than it might seem: it changes how expensive attention is, and it changes how much temporal structure the backbone actually gets to see.

\paragraph{Backbone architecture.}
Every backbone stacks $N{=}6$ layers with $d{=}256$. Each layer is one of two types.

The \textbf{Bidirectional Mamba Block (BiMamba)} runs two independent selective SSMs, one forward and one on the time-reversed input, and merges their outputs:
\begin{align}
\mathbf{H} &= \mathrm{LayerNorm}(\mathbf{X}), \label{eq:bimamba-norm}\\
\overrightarrow{\mathbf{Y}} &= \mathrm{Mamba}_{\text{fwd}}(\mathbf{H}), \label{eq:bimamba-fwd}\\
\overleftarrow{\mathbf{Y}} &= \mathrm{Flip}\!\bigl(\mathrm{Mamba}_{\text{bwd}}(\mathrm{Flip}(\mathbf{H}))\bigr), \label{eq:bimamba-bwd}\\
\mathbf{X}' &= \mathbf{X} + \mathbf{W}_{\text{proj}}\,[\overrightarrow{\mathbf{Y}};\,\overleftarrow{\mathbf{Y}}], \label{eq:bimamba-proj}
\end{align}
where $[\,\cdot\,;\,\cdot\,]$ is feature concatenation and $\mathbf{W}_{\text{proj}}\in\mathbb{R}^{d\times 2d}$. Standard Mamba is causal; bidirectionality is necessary for classification where the model needs access to the full input. We use $d_{\text{state}}{=}32$, $d_{\text{conv}}{=}4$, expansion factor $E{=}2$. The block is $\mathcal{O}(L)$ in time and memory.

The \textbf{Self-Attention Block} is a pre-norm Transformer layer with multi-head attention ($h{=}4$) and a GELU FFN:
\begin{align}
\mathbf{H} &= \mathrm{LayerNorm}(\mathbf{X}), \label{eq:attn-norm}\\
\mathbf{A} &= \mathbf{X} + \mathrm{MHA}(\mathbf{H},\mathbf{H},\mathbf{H}), \label{eq:attn-mha}\\
\mathbf{X}' &= \mathbf{A} + \mathrm{FFN}\!\bigl(\mathrm{LayerNorm}(\mathbf{A})\bigr). \label{eq:attn-ffn}
\end{align}
The FFN width is not $4d$; it is set by parameter matching (Section~\ref{sec:param-match}). At our model scale, six layers of self-attention on 30,000 tokens exceeds 48\,GB of GPU memory, which is what makes the backbone choice consequential at long sequence lengths.

\paragraph{Attention ratio.}
We test three points along this axis:

\begin{itemize}[leftmargin=0pt,itemindent=0pt,labelsep=4pt,itemsep=2pt]
    \item \textbf{Pure Mamba.} Six BiMamba layers. Zero attention. $\mathcal{O}(L)$.
    \item \textbf{HELIX.} Five BiMamba layers with one attention layer inserted at position 3 (middle of the stack). Three Mamba layers build local features, one attention layer provides global interaction, two final Mamba layers refine the result. One layer of $\mathcal{O}(L^2)$ on an otherwise linear backbone.
    \item \textbf{Pure Attention.} Six attention layers. $\mathcal{O}(L^2)$.
\end{itemize}

In effect, these three points (0, 1, and 6 attention layers at matched parameter budget) give us a coarse ablation over the amount of attention in the model. We place the single attention layer at position 3 (mid-stack) so that the first three Mamba layers build local features before global interaction, and the final two Mamba layers can refine the globally informed representation. Placing attention too early would force it to operate on raw, unprocessed features where global patterns are not yet visible; placing it too late would leave the final layers with no room to refine the globally informed representation. Mid-stack placement follows the same reasoning used in hybrid language models such as Jamba~\cite{lieber2024jambahybridtransformermambalanguage}, where global mixing is inserted after local structure has been established.

Crossing three backbones with two frontends gives six model variants. The goal is not to crown a single winner. We want to see whether the ranking between these variants actually shifts when the task or sequence length changes, and if so, to understand why.

\subsection{Parameter Matching}
\label{sec:param-match}

Fair comparison requires that all models have the same capacity; otherwise any accuracy difference could just reflect having more parameters. We enforce this at the layer level. The BiMamba block serves as the reference with $P_{\text{mamba}}$ parameters. For each attention block, we solve for the FFN width that equalizes its count:
\begin{equation}
d_{\text{ffn}} = \left\lfloor \frac{P_{\text{mamba}} - P_{\text{MHA}} - P_{\text{norms}}}{2d + 1} \right\rfloor,
\label{eq:ffn-dim}
\end{equation}
where $P_{\text{MHA}} = 4d^2 + 4d$ covers the projection matrices and biases, and $P_{\text{norms}} = 4d$ for two LayerNorms. This is computed once at construction. Every layer, regardless of type, has nearly identical parameter count. All six variants land at ${\sim}8.3$M total. A 10\% accuracy gap between two variants is architectural, not a capacity artifact.

\subsection{Classification Head and Selective Pooling}

After the last backbone layer, LayerNorm and mean pooling aggregate the token sequence into a single vector:
\begin{equation}
\hat{\mathbf{y}} = \mathbf{W}_{\text{cls}}\;\frac{1}{K}\sum_{i=1}^{K}\mathbf{x}_i^{(N)}.
\label{eq:cls-head}
\end{equation}
By default $K{=}L$ (all tokens). For our long-sequence experiments (Concat Speech Commands, Section~\ref{sec:experimental-setup}), where multiple clips are concatenated and the label belongs to the first clip, we pool only the first 100 tokens ($K{=}100$). We chose this deliberately: it converts the task into a test of long-range information retention. The model processes the full $L$-token sequence causally through every backbone layer, but the gradient signal flows only through the first 100 output positions. A backbone that lets early information decay will fail regardless of its overall representational power. This setup isolates the memory axis of the comparison more cleanly than global pooling would, because global pooling allows a model to succeed by attending to any informative region rather than requiring it to preserve a specific temporal location.

\subsection{Training Protocol}

All variants use the same training setup to prevent confounding. AdamW~\cite{loshchilov2019decoupledweightdecayregularization} with learning rate $3{\times}10^{-4}$, weight decay 0.05, cosine annealing to $10^{-6}$. Gradient clipping at norm 1.0, which we find necessary: the SSM variants occasionally diverge without it. Mixed-precision (FP16) where GPU memory requires it.

For augmentation we apply random circular time shifts ($\pm$0.5\,s), amplitude scaling ($[0.8, 1.2]$), Gaussian noise ($\sigma{=}0.005$), and mixup~\cite{zhang2018mixupempiricalriskminimization} ($\alpha{=}0.3$). The mixup mattered most on small datasets like ESC-50.

Datasets with predefined folds (ESC-50: 5-fold; UrbanSound8K: 10-fold) are reported as mean $\pm$ std of best test accuracy across folds. Single-split datasets report best test accuracy. Batch size 32, 100 epochs, unless compute limits intervened.

\section{Experimental Setup}
\label{sec:experimental-setup}

With the framework defined, we now describe the datasets and compute environment. We chose each experiment group to expose a specific interaction between design axes, the kind that would go unnoticed if the axes were studied one at a time.

\subsection{Datasets}

\paragraph{Short-duration classification.}
At these sequence lengths attention is still computationally feasible, so any differences between architectures come down to inductive bias rather than compute constraints.

\textbf{ESC-50}~\cite{10.1145/2733373.2806390}: 2,000 environmental sound clips at 5\,s each (50 classes: rain, dog bark, chainsaw, etc.). 5-fold cross-validation with predefined folds. At 16\,kHz the raw path produces $L{=}500$ tokens; the spectrogram path produces $L'{\approx}80$. This is the simplest benchmark and the one where we expect the least benefit from attention, since most environmental sounds have stationary or repetitive structure within a 5\,s window.

\textbf{UrbanSound8K}~\cite{Salamon:UrbanSound:ACMMM:14}: 8,732 urban sound clips at 4\,s each (10 classes). 10-fold cross-validation. Similar scale to ESC-50 but fewer, broader classes.

\textbf{Speech Commands v2}~\cite{warden2018speechcommandsdatasetlimitedvocabulary}: ${\sim}$105k spoken single-word commands at 1\,s each (35 classes: ``yes,'' ``no,'' ``left,'' ``stop,'' etc.). Single official train/test split. The raw path yields just $L{=}100$ tokens per clip. This is where we expect the input representation axis to matter least, since even raw waveforms produce short sequences.

\paragraph{Long-sequence memory.}
To test whether backbone differences grow with sequence length, we construct a synthetic long-range task from Speech Commands.

\textbf{Concat Speech Commands}: We concatenate $n{=}10$ random clips from the same split into a single 10\,s input ($L{=}1{,}000$ tokens on the raw path). The label is the class of the \emph{first} clip. The model must identify the first command and retain that information while processing nine subsequent, unrelated commands. We restrict pooling to the first 100 tokens ($K{=}100$). All architectures see the same long input. The only question is which backbone best preserves the early information.

\paragraph{Speaker identification at scale.}
Speaker ID is a natural testbed for the sequence-length axis. Longer audio contains more speaker-discriminative information, but only if the model can actually aggregate fine-grained vocal characteristics across the full input. Whether it can depends on exactly the design choices we are varying. We use two corpora at different scales:

\textbf{LibriSpeech}~\cite{7178964} (train-clean-100): 100 hours of read English speech, 921 speakers. Utterances from the same speaker are stitched together to create clips of a target duration. We split utterances per speaker (80\% train, 20\% val) so all speakers appear in both splits (closed-set identification). We evaluate at 30\,s ($L{=}3{,}000$ tokens).

\textbf{VoxPopuli}~\cite{wang2021voxpopulilargescalemultilingualspeech} (English): European Parliament recordings, 1,313 speakers. Segments are naturally longer than LibriSpeech (tens of seconds), requiring fewer stitches per clip. We evaluate at 300\,s (5 minutes, $L{=}30{,}000$ tokens). This is the regime we were most interested in. The raw path produces 30,000 tokens, which is already too many for pure attention. Spectrograms bring the count down to a few thousand, but at the cost of the fine acoustic detail that distinguishes one speaker from another. So the question becomes: can any backbone actually aggregate 30,000 tokens of speaker information without either running out of memory or losing the signal?

\subsection{Compute and Scope}

All experiments run on NVIDIA RTX 6000 Pro GPUs. ESC-50 and UrbanSound8K complete in under 5 hours per variant. Speech Commands takes roughly 12 hours. The 5-minute speaker ID experiments take 40+ hours per run. We use Weights \& Biases for tracking.

Table~\ref{tab:scaling-cost} summarizes the computational feasibility of each backbone as sequence length grows. The asymptotic gap is well known ($\mathcal{O}(L)$ vs.\ $\mathcal{O}(L^2)$), but what matters for us is the practical consequence: Pure Attention drops out of the comparison at exactly the sequence lengths we care about most.

\begin{table}[h]
\centering
\caption{Computational feasibility at ${\sim}$8.3M parameters on a single RTX 6000 Pro (48\,GB). $\checkmark$ = trains normally; \textit{slow} = feasible but $>$2$\times$ wall time vs.\ Pure Mamba; OOM = out of memory.}
\label{tab:scaling-cost}
\small
\begin{tabular}{@{}lccc@{}}
\toprule
Tokens & Pure Mamba & HELIX & Pure Attention \\
\midrule
500 (5\,s) & $\checkmark$ & $\checkmark$ & $\checkmark$ \\
3{,}000 (30\,s) & $\checkmark$ & $\checkmark$ & \textit{slow} \\
30{,}000 (5\,min) & $\checkmark$ & $\checkmark$ & OOM \\
\bottomrule
\end{tabular}
\end{table}

These costs also constrain the experimental design. A single long-sequence run costs over 65 GPU-hours, making exhaustive sweeps over attention placement or layer count impractical. We therefore prioritize \emph{breadth across datasets and sequence regimes} over fine-grained ablations of any one hyperparameter. The three backbone variants (0, 1, and 6 attention layers) already span the extremes of the attention ratio axis at matched capacity, which we believe is enough to show that these design choices are not independent. Where runs were terminated early by compute limits, we verify convergence from learning curves and report the epoch count explicitly.

\section{Results}

\begin{figure*}[t]
\centering
\includegraphics[width=\textwidth]{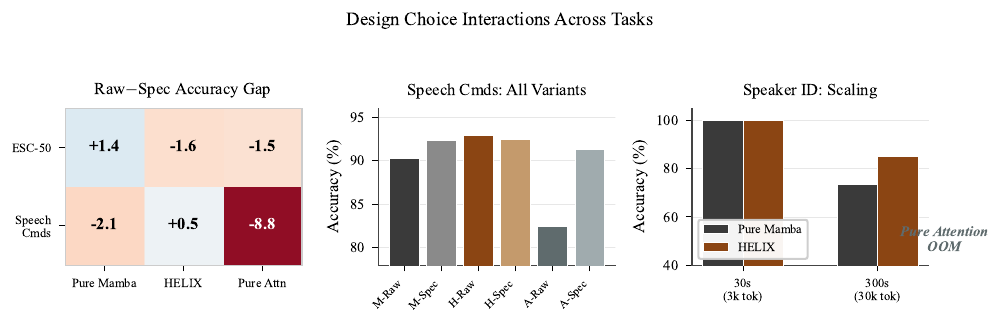}
\caption{Overview of design choice interactions. \textbf{Left:} The raw-vs-spectrogram accuracy gap (positive = raw better) depends on backbone. Pure Mamba prefers raw on ESC-50 but not Speech Commands; Pure Attention strongly prefers spectrograms. \textbf{Center:} All six Speech Commands variants. \textbf{Right:} As sequence length scales from 3k to 30k tokens, the HELIX advantage grows from 0 to 11.5 points; Pure Attention cannot run at 30k tokens.}
\label{fig:hero}
\end{figure*}

We present results grouped by which interaction they reveal. All accuracy numbers are best test accuracy (mean $\pm$ std across folds where applicable). Runs marked with $^\dagger$ were terminated early by compute time limits but had converged or nearly converged based on learning curves. Figure~\ref{fig:hero} provides an overview of how the three design axes interact across tasks.

\subsection{Short-Duration Classification}

\begin{figure}[h]
\centering
\includegraphics[width=\columnwidth]{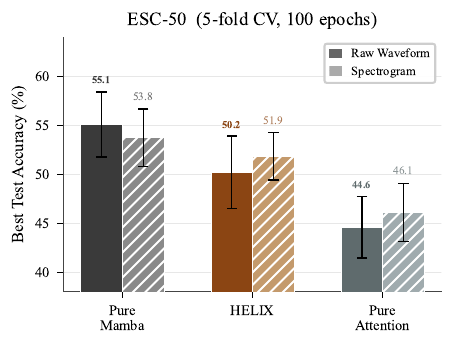}
\caption{ESC-50 results. Solid bars: raw waveform; hatched bars: spectrogram. Error bars show $\pm$1 std across 5 folds. Pure Mamba dominates; input preference flips between backbones.}
\label{fig:esc50}
\end{figure}

\begin{figure}[h]
\centering
\includegraphics[width=\columnwidth]{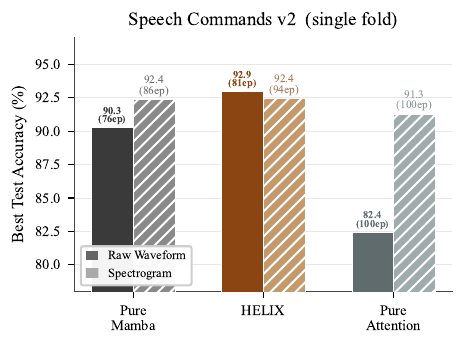}
\caption{Speech Commands v2 results. Epoch counts shown above bars (runs that hit compute limits did not reach 100 epochs). Pure Attention raw collapses; HELIX raw leads despite fewer epochs.}
\label{fig:speechcommands}
\end{figure}

Table~\ref{tab:esc50} and Figure~\ref{fig:esc50} show ESC-50 results across all six variants. Table~\ref{tab:speechcommands} and Figure~\ref{fig:speechcommands} show Speech Commands.

\begin{table}[h]
\centering
\caption{ESC-50 environmental sound classification (5-fold CV, 100 epochs). Best accuracy per column in \textbf{bold}.}
\label{tab:esc50}
\small
\begin{tabular}{@{}lcc@{}}
\toprule
Architecture & Raw & Spectrogram \\
\midrule
Pure Mamba & \textbf{55.10} $\pm$ 3.33 & \textbf{53.75} $\pm$ 2.92 \\
HELIX & 50.20 $\pm$ 3.71 & 51.85 $\pm$ 2.41 \\
Pure Attention & 44.60 $\pm$ 3.16 & 46.10 $\pm$ 2.97 \\
\bottomrule
\end{tabular}
\end{table}

\begin{table}[h]
\centering
\caption{Speech Commands v2 (single split). $^\dagger$Terminated before 100 epochs due to compute limits. Ranking is consistent with trends observed in validation curves across partial runs. Best per column in \textbf{bold}.}
\label{tab:speechcommands}
\small
\begin{tabular}{@{}lcc@{}}
\toprule
Architecture & Raw & Spectrogram \\
\midrule
Pure Mamba & 90.27$^\dagger$ (76 ep) & 92.36$^\dagger$ (86 ep) \\
HELIX & \textbf{92.94}$^\dagger$ (81 ep) & \textbf{92.44}$^\dagger$ (94 ep) \\
Pure Attention & 82.43 (100 ep) & 91.27 (100 ep) \\
\bottomrule
\end{tabular}
\end{table}

Two things surprised us here.

\textbf{The input representation preference depends on the backbone.} On ESC-50, Pure Mamba prefers raw waveforms (55.10 vs.\ 53.75), but Pure Attention prefers spectrograms (46.10 vs.\ 44.60). This is not a small effect. The best raw model (Mamba, 55.10) beats the best spectrogram model (also Mamba, 53.75), but if someone tested only Pure Attention they would conclude the opposite: that spectrograms are the better representation. The takeaway is not that raw is better or spectrograms are better; it depends on what you pair them with.

\textbf{Attention hurts on short stationary audio.} On ESC-50, where clips are 5\,s of environmental sound with repetitive structure, Pure Mamba outperforms HELIX by 5 points and Pure Attention by over 10. Put simply, there is nothing for attention to attend \emph{to} that the SSM's recurrence does not already capture. The global interaction layer replaces one BiMamba layer that was doing useful local processing, and on this task that trade-off is not worth it.

\begin{figure*}[t]
\centering
\includegraphics[width=\textwidth]{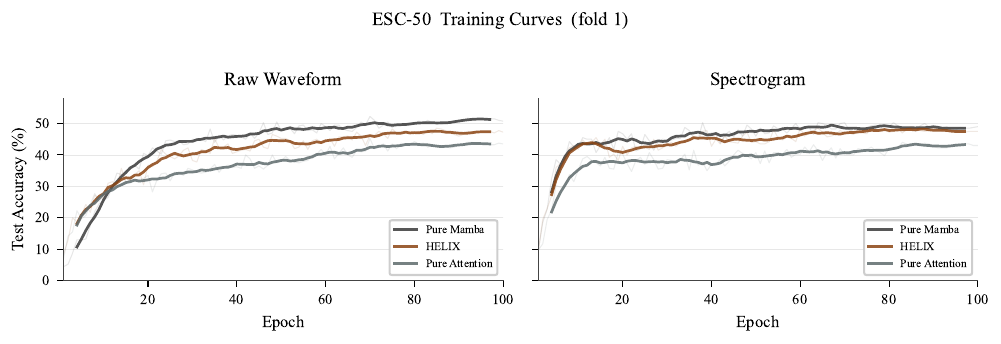}
\caption{ESC-50 training curves (fold 1). Raw waveform models (left) learn slower but Pure Mamba eventually surpasses all spectrogram variants (right). Smoothed with a 7-epoch moving average; raw values shown underneath.}
\label{fig:esc50-curves}
\end{figure*}

Speech Commands tells a different story. HELIX raw (92.94\%) overtakes Pure Mamba raw (90.27\%) by 2.7 points. Spoken commands have more temporal structure than environmental sounds: the difference between ``left'' and ``right'' is in the precise ordering of phonemes, not just their spectral content. One layer of attention helps the model align these temporal cues. Pure Attention raw collapses to 82.43\%, nearly 10 points behind. We think the issue is that six layers of quadratic attention on raw tokens has no inductive bias for local continuity. Audio is overwhelmingly local in structure, and attention spreads its capacity across all pairwise interactions without prioritizing nearby ones. At our model scale this seems to waste parameters and destabilize training. On spectrograms, where the sequence is only ${\sim}10$ tokens and local structure has already been captured by the STFT, Pure Attention recovers to 91.27\%.

\begin{figure*}[t]
\centering
\includegraphics[width=\textwidth]{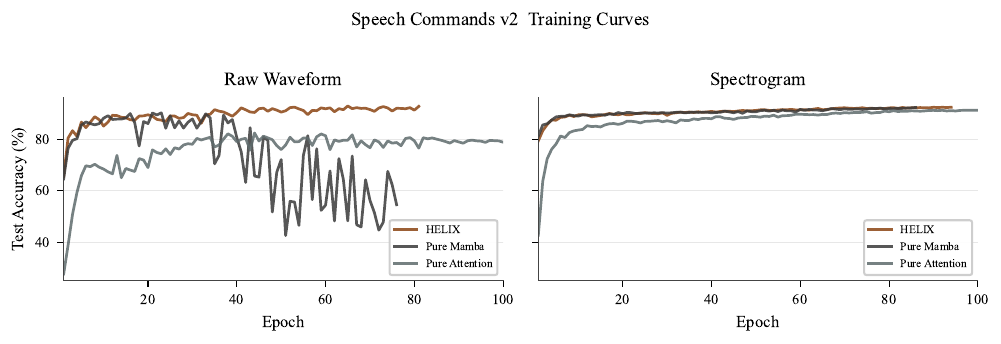}
\caption{Speech Commands training curves. \textbf{Left:} On raw waveforms, Pure Mamba raw destabilizes after epoch 30 and never recovers, while HELIX raw climbs steadily to 92.9\%. Pure Attention raw plateaus at ${\sim}82$\%. \textbf{Right:} Spectrograms stabilize all architectures; the gap between variants shrinks to ${\sim}1$\%.}
\label{fig:sc-curves}
\end{figure*}

\subsection{Long-Sequence Memory}

\begin{table}[h]
\centering
\caption{Concatenated Speech Commands ($n{=}10$ clips, $L{=}1{,}000$ tokens). Label from first clip; pooling restricted to first 100 tokens. Both runs crashed at 24 epochs.}
\label{tab:longseq}
\small
\begin{tabular}{@{}lcc@{}}
\toprule
Architecture & Best Acc (\%) & Final Acc (\%) \\
\midrule
HELIX Raw & \textbf{91.31} & 90.44 \\
Pure Mamba Raw & 89.81 & 89.36 \\
\bottomrule
\end{tabular}
\end{table}

\begin{figure}[!t]
\centering
\includegraphics[width=\columnwidth]{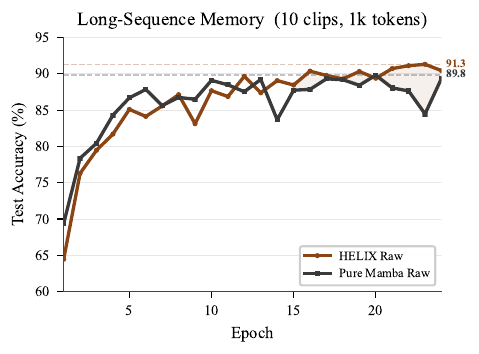}
\caption{Long-sequence training curves. HELIX separates from Pure Mamba after epoch 15 and maintains the gap. Both runs were still improving when compute ran out.}
\label{fig:longseq}
\end{figure}

With the sequence at 1,000 tokens and the label depending only on the first clip, HELIX leads Pure Mamba by 1.5 points after 24 epochs (Table~\ref{tab:longseq}, Figure~\ref{fig:longseq}). We want to be upfront about the limitation: both runs were terminated by compute constraints and were still improving, so this gap is preliminary. We include it because the direction is consistent with the longer-scale speaker ID results below, and because the training curves (Figure~\ref{fig:longseq}) show a clean separation after epoch 15 that widens steadily rather than fluctuating. But on its own, a 1.5-point gap at 24 epochs is suggestive, not conclusive.

\subsection{Speaker Identification at Scale}

\begin{table}[h]
\centering
\caption{Speaker identification on VoxPopuli (English), 300\,s clips (1,313 speakers, $L{=}30{,}000$ tokens). Pure Attention could not run due to OOM.}
\label{tab:scaling}
\small
\begin{tabular}{@{}lcc@{}}
\toprule
Architecture & Best Acc (\%) \\
\midrule
Pure Mamba Raw & 73.55 \\
HELIX Raw & \textbf{85.07} \\
\bottomrule
\end{tabular}
\end{table}

\begin{figure}[!ht]
\centering
\includegraphics[width=\columnwidth]{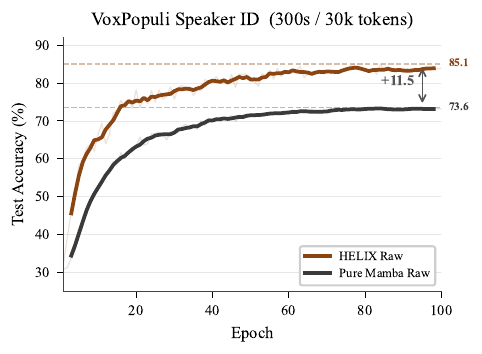}
\caption{VoxPopuli 300\,s training curves. HELIX pulls ahead early and reaches 85\% while Pure Mamba plateaus near 73\%. Dotted lines mark best accuracy. This is the largest gap in our study.}
\label{fig:voxpopuli}
\end{figure}

Table~\ref{tab:scaling} and Figure~\ref{fig:voxpopuli} show our longest experiment. VoxPopuli at 300\,s (30,000 tokens, 1,313 speakers) is the most demanding setting in our study, and frankly the one we learned the most from. This is where the three design axes start to pull in different directions:

\textbf{Backbone:} HELIX (85.07\%) outperforms Pure Mamba (73.55\%) by 11.5 points. We did not expect a single attention layer to account for this large a gap at equal parameter count. The SSM backbone efficiently processes the long sequence, but the attention bottleneck at position 3 provides a global summary that helps the final layers aggregate speaker characteristics spread across 5 minutes of speech. Pure Attention cannot run at all at this sequence length: six layers of self-attention on 30,000 tokens exceeds GPU memory on the RTX 6000 Pro.

\textbf{Input representation:} We run the raw waveform path here because the spectrogram path, while computationally feasible ($L'{\approx}3{,}750$ tokens at 300\,s), discards the fine acoustic detail (pitch micro-variations, vocal tract resonances) that distinguishes speakers. The speaker ID task at this scale rewards preserving the full signal.

\textbf{The entanglement:} In practice, these choices interact in ways that are hard to predict from any single axis. HELIX + raw waveform at 300\,s achieves 85\%. Pure Mamba + raw at the same length gets 73.5\%. Pure Attention + raw cannot run. Pure Attention + spectrogram could run in principle (3,750 tokens is feasible), but spectrogram compression discards the exact information the task needs. You cannot get to 85\% by optimizing along any single axis; the result depends on getting all three right together.

\section{Analysis and Conclusion}

The central finding across our experiments is that input representation, backbone family, and attention ratio are not independent design choices. The frontend controls both what information the model sees and how long the resulting sequence is. The backbone then has to handle that sequence length, which may or may not be feasible depending on its complexity class. And the attention ratio affects whether the model can coordinate information across temporal distances that exceed what recurrence alone can maintain. These dependencies mean that changing any one axis can shift the ranking on the others, often in ways we found difficult to anticipate. The parameter-matched setup lets us observe this directly rather than piecing it together from separate papers with different scales and budgets.

\paragraph{Why does a single attention layer help at scale?}
We do not have a mechanistic answer to this question. We did not probe internal representations or ablate attention position, so what follows is a hypothesis consistent with the results, not a tested explanation.

The HELIX advantage over Pure Mamba is near zero on short stationary audio, moderate on temporally structured speech, and largest (11.5 points) at 30,000 tokens. Why would it scale this way? One candidate explanation is information decay in recurrence. In a pure SSM, information from token $t$ must survive $L - t$ steps to influence the output. Mamba's selective gating slows this decay but does not eliminate it; on short sequences the loss is negligible, but over 30,000 steps even small per-step attenuation compounds. A single attention layer at mid-stack could act as a global synchronization point, letting every token read from every other token once and creating a compressed summary that subsequent Mamba layers propagate more efficiently than raw recurrent state. The training curves are at least consistent with this: on VoxPopuli 300\,s (Figure~\ref{fig:voxpopuli}), HELIX separates from Pure Mamba early and the gap widens steadily, which makes us think the advantage is structural rather than an optimization artifact. But we want to be clear that probing experiments (e.g., attention maps, token-level retention analysis, or position ablations) would be needed to confirm this account.

\paragraph{HELIX as a scaling strategy.}
The VoxPopuli result reframes the hybrid not as a better architecture in general, but as a scaling strategy for a specific regime. On short sequences where recurrence alone carries all necessary information (ESC-50, LibriSpeech 30\,s), the attention layer is a net cost: it replaces a BiMamba block that was doing useful local processing. The picture changes at longer scales. At 30,000 tokens, full attention is computationally infeasible, but a single attention bottleneck provides just enough global interaction at minimal quadratic cost. If we had to distill a practical guideline from these results, it would be: let recurrence handle most of the sequence, and spend a small attention budget at the point where the model needs to pull together information that is spread across a long time span.

\paragraph{Implications for audio architecture design.}
The broader conclusion is more specific than ``raw is better'' or ``SSMs beat transformers.'' Raw waveforms help most when the task depends on fine acoustic detail and the backbone can handle the resulting token count, but spectrogram compression is the better choice when optimization stability and sequence shortening matter more than preserving every cue. On the backbone side, pure Mamba is hard to beat when local sequential structure dominates. HELIX pulls ahead as temporal dependencies grow longer. We think the main takeaway for practitioners is to treat representation and sequence model as a joint design problem. Conclusions we drew from short-sequence experiments did not hold at longer scales, and we suspect the same is true for results in the broader literature.

\paragraph{Limitations and scope.}
Several runs were compute-constrained (65+ GPU-hours per long-sequence experiment), and we report epoch counts explicitly where training did not complete. We evaluate one hybrid placement (mid-stack) and three points on the attention ratio axis (0, 1, 6 layers). Finer sweeps would strengthen the analysis, but even with just three points the effects are clear and consistent. We also only test classification; whether the same interactions show up in generation or dense prediction is an open question we would like to revisit. Even so, the overall picture is consistent across our experiments. Short stationary audio favors efficient recurrence. Moderately structured speech benefits from a small amount of global interaction. And at the longest scales we tested, hybrids that spend attention sparingly outperform both extremes by a substantial margin.

\section*{Acknowledgements}
We thank the Safety and Alignment Research India community for providing access to RTX 6000 Pro compute that made the full-scale experiments in this work possible.

\bibliography{example_paper}
\bibliographystyle{icml2025}

\end{document}